\newcommand{\be}{\begin{eqnarray}}
\newcommand{\ee}{\end{eqnarray}}
\newcommand{\dia}{\!\!\!\!\!\!\not\,\,\,}

\def\mathnew{\mathsurround=0pt}
\def\simov#1#2{\lower .5pt\vbox{\baselineskip0pt
    \lineskip-.5pt\ialign{$\mathnew#1\hfil##\hfil$\crcr#2\crcr\sim\crcr}}}
\def\simgreat{\mathrel{\mathpalette\simov >}}

\documentclass{svmult}

\usepackage{makeidx}     
\usepackage{graphicx}    
\usepackage{multicol}    

\makeindex             

\begin{document}

\title*{Electroweak baryogenesis and primordial hypermagnetic fields}
\author{Gabriella Piccinelli\inst{1}\and
Alejandro Ayala\inst{2}}
\institute{Centro Tecnol\'ogico Arag\'on,
         Universidad Nacional Aut\'onoma de M\'exico,
         Av. Rancho Seco S/N, Bosques de Arag\'on, Nezahualc\'oyotl,
         Estado de M\'exico 57130, M\'exico\\ 
\texttt{gabriela@astroscu.unam.mx}
\and Instituto de Ciencias Nucleares,
           Universidad Nacional Aut\'onoma de M\'exico,
           Apartado Postal 70-543, M\'exico Distrito Federal 04510,
           M\'exico
\texttt{ayala@nuclecu.unam.mx}}
%
%
\maketitle

The origin of the matter-antimatter asymmetry of the universe remains
one of the outstanding questions yet to be answered by modern
cosmology and also one of only a handful of problems where the
need of a larger number of degrees of freedom than those contained in
the standard model (SM) is better 
illustrated. An appealing scenario for the generation of baryon number
is the electroweak phase transition that took place when the
temperature of the universe was about 100 GeV. Though in the minimal version of the SM, and without considering the interaction of the SM particles with additional degrees of freedom, this scenario has been ruled out given the current bounds for the  Higgs mass, this still remains an open possibility in supersymmetric
extensions of the SM. In recent years it has also been realized that
large scale magnetic fields could be of primordial origin. A natural
question is what effect, if any, these fields could have played during
the electroweak phase transition in connection to the generation of
baryon number. Prior to the electroweak symmetry breaking, the
magnetic modes able to propagate for large distances belonged to the
$U(1)$ group of hypercharge and hence receive the name of \emph{
hypermagnetic} fields. In this contribution, we summarize recent work
aimed to explore the effects that these fields could have introduced
during a first order electroweak phase transition. In particular, we
show how these fields induce a CP asymmetric scattering of fermions
off the true vacuum bubbles nucleated during the phase transition. The
segregated axial charge acts as a seed for the generation of baryon
number. We conclude by mentioning possible research venues to further
explore the effects of large scale magnetic fields for the generation
of the baryon asymmetry. 

\section{Introduction}\label{I}

The standard model (SM) of electroweak interactions meets all the
requirements --commonly referred to as Sakharov
conditions~\cite{Sakharov}-- to 
generate a baryon asymmetry during the electroweak phase transition
(EWPT), provided that this last be of first order. However, it is also
well known that neither the amount of CP violation within the
minimal SM nor the strength of the EWPT are enough to generate a
sizable baryon number~\cite{{Gavela},{Kajantie}}. Supersymmetric
extensions of the SM, with a richer particle content, contain new
sources of CP violation~\cite{CKN}. They also allow a stronger first
order phase transition~\cite{Carena}. In spite of these improvements
with respect to the SM, the minimal supersymmetric model (MSSM) is
severely constrained from experimental bounds on the chargino
properties~\cite{Cline} leaving only a small corner of parameter space
for the MSSM as a viable candidate for baryogenesis. Further
possibilities to accommodate an explanation for the generation of
baryon number during the EWPT include non-minimal supersymmetric
models which, nonetheless, all share the unappealing feature of
containing an even larger set of parameters than the already extensive
number contained in the MSSM. 

Though it might appear tempting to abandon the idea of electroweak
baryogenesis (EWB) given the above difficulties, in recent years this
possibility has been revisited due to the observation that magnetic
fields are able to generate a stronger first order
EWPT~\cite{{Giovannini},{Elmfors},{Giovannini2}}. The situation is  
similar to what happens to a type I superconductor where an external
magnetic field modifies the nature of the superconducting phase
transition, changing it from second to first order due to the Meissner
effect. 

In spite of this observation, it has also been
realized that the sphaleron bound becomes more restrictive due to the
interaction between the sphaleron's magnetic dipole moment and the
external field~\cite{Comelli}. Nevertheless, these arguments are
either classical or resort to perturbation theory to lowest
order. However, in the absence of magnetic fields, it is well known
that the phase transition picture is influenced by non-perturbative
effects cast in terms of the resummation of certain classes of
diagrams~\cite{Carrington} and it might also be expected that the same
is true in the presence of magnetic fields. The situation with regards
to the strength of the phase transition in the presence of magnetic
fields is thus far from being settled and requires further research.

However, the influence on the magnetic fields on the enhancement of CP
violation has received much less attention. In a series of recent
papers~\cite{Ayala1, Ayala2, Ayala3}, it has been shown that the
external field is able to produce an axially asymmetric scattering of
fermions off first order phase transition bubbles during the
EWPT. This CP violating reflection is 
due to the chiral nature of the couplings of right- and left-handed modes
with the external field in the symmetric phase . This mechanism
produces an axial charge segregation which can then be transported in
the symmetric phase where sphaleron induced transitions can convert it
into baryon number~\cite{Nelson}. The main purpose of our work is the
description of the mechanism for the generation of this axial charge
segregation.  

As was briefly presented in this introduction, the process of
baryogenesis involves several physical ingredients which all deserve
to be addressed in order to  cover the entire topic; here we will
concentrate on those aspects related to hypermagnetic fields, which are
also one of the most recent and less explored parts of this field. For
other aspects on the subject, we refer the reader to recent
reviews~\cite{Trodden, Riotto, Rubakov}.  

The work is organized as follows: in Sec.~\ref{II} we present the
basic framework of electroweak baryogenesis. Sec.~\ref{III} is
devoted to generalities of hypermagnetic fields and phase transitions. In
Sec.~\ref{IV} we summarize the current ideas for the 
origin of large scale magnetic fields as well as the experimental
bounds on their strength set by different observations. In
Sec.~\ref{V} we describe the mechanism whereby the asymmetric
reflection of fermions off first order EWPT bubbles in the presence of
external magnetic fields leads to an axial charge segregation that can
then be converted into baryon number in the symmetric phase. Finally,
in Sec.~\ref{VI} we summarize and give a brief account of some possible
research venues to further explore the influence of primordial
magnetic fields in the generation of the baryon asymmetry of the
universe.  

\section{Electroweak baryogenesis}\label{II}

\subsection{Baryogenesis}\label{II.1}

The theory of baryogenesis is an intent to explain the existence of
matter in the Universe. As Cohen, Kaplan and Nelson~\cite{CKN} put it:
why is there something rather than nothing? 

From the point of view of elementary particle physics, there is a
symmetry between particles and antiparticles which suggests that there
should be an overall balance between the amount of matter and
antimatter in the universe. However the observed universe is composed
almost entirely of matter, with no traces of present or primordial
antimatter (see. e.g.~\cite{Kolb} pp 158--159, or~\cite{Stecker} for a
recent review). On the other hand, from the cosmological approach,
there is also some evidence that some ingredients are missing. In
the hot early epoch of the universe evolution, one expects to have
particles and antiparticles in thermal equilibrium with radiation;
particle/antiparticle pairs would then annihilate each other until
their annihilation rate becomes smaller than the rate of expansion of
the universe. The remaining density of
all the species  can thus be estimated and it comes to be only
a very small fraction of the closure density of the universe.  
In this way, if we do not wish to postulate that the universe was just
born with ad hoc asymmetric initial conditions, we must find a
mechanism to generate a net baryon number ($B = n_b - n_{\bar b}$).  

In 1966, Sakharov~\cite{Sakharov} laid out the conditions for the
development of a net excess of baryons over antibaryons:  
(1) Existence of interactions that violate baryon number; (2) C and CP
violation and (3) departure from thermal equilibrium.  
(The implications of each one of these criteria is discussed in many
places, see e.g. Ref.~\cite{Trodden}).  

It must be stressed however that some scenarios (possibly exotic) have
been recently proposed where one of these conditions is not achieved
(see e.g. Ref.~\cite{Dolgov} and references therein). 

It is important to notice that a successful model for baryon generation has to put together two ingredients:

1) the generation of a baryon asymmetry 

2) its preservation

\noindent In the following subsection, we will review the necessary conditions for both situations in the framework of EWB.

\subsection{Electroweak baryogenesis}\label{II.2}

\subsubsection{Sakharov conditions in the standard model}

The sphaleron (the name is based on the classical greek adjective
meaning ``ready to fall")~\cite{Klink} is an static and unstable
solution of the field equations of the EW model, corresponding to the
top of the energy barrier between two topologically distinct
vacua. Transitions between these vacua are associated with the violation
of baryon (B) and lepton number (L)~\cite{'tHooft}, in the combination
B + L, with leptons and baryons produced with the same rate (i.e. B - L
conservation).  
For this reason, they can either induce baryogenesis, or be a
mechanism for washing out the previously created baryon asymmetry. It
is therefore important to define the epoch at which the sphaleron
transitions fall out of thermal equilibrium. 

As we mentioned, C and CP violation are present in the SM but are too
tiny to be at the origin of the present baryon asymmetry~\cite{Gavela,
  Benreu}. The generation of a sizable CP violation in the SM is the
central part of this work and we will return to it later. 
 
For the out of equilibrium requirement, we rely on the phase transition (PT). 
This is the only possible source of departure from thermal
equilibrium since, at the electroweak scale, the rate of expansion of
the universe
is small compared to the rate of baryon number violating processes. 
But the PT is efficient in producing out-of-equilibrium conditions
only if it is strongly first order~\cite{Kuzmin}, i.e. if the Higgs
field --which is the order parameter in this case-- undergoes a
discontinuous change. In effect, in a first order PT, the conversion
from one phase to another happens through nucleation and propagation
of the true vacuum bubbles. The region separating both phases is
called the wall. As the bubble wall sweeps a region in space, the order
parameter changes rapidly, leading to a departure from thermal
equilibrium. 

\subsubsection{Preservation of the baryon asymmetry}

In order to freeze out the produced baryon number, the rate
of fermion number non conservation in the broken phase, at
temperatures below the bubble nucleation temperature, must be smaller
than the rate of expansion of the universe.  

The rate, per unit volume, of baryon number violating events, depends
at low temperatures $T < T_c$ (more precisely for $M_W << T << M_W/
\alpha_W$) on the sphaleron energy: 

\be
\Gamma = \mu \left(\frac{M_W}{\alpha_W T}\right)^3 M_W^4 \exp\left(\frac{E_{\mbox{\tiny{sph}}}(T)}{T}\right)
\label{sphal1}\, ,
\ee
with $\mu$ a dimensionless constant and $E_{\mbox{\tiny{sph}}} \sim
M_W(T) / \alpha_W$ ($\alpha_W = g^2/ 4 \pi$, with $g$ the $SU(2)_L$
gauge coupling). Comparing this rate with the rate of expansion of the
universe $H \sim g_*^{1/2} T^2 / m_{Pl}$~\cite{Kolb} pp 60--65 ($g_*$ is the
effective number of degrees of freedom and $m_{Pl}$ is the Planck
mass), the following bound is found~\cite{Shap1}:  

\be
E_{\mbox{\tiny{sph}}}(T_{\mbox{\tiny{nucl}}}) / T_{\mbox{\tiny{nucl}}}
> A \ \ ; \ \  A \simeq 35 - 45   
\label{sphal2}\, .
\ee
Here, we assume that the major wash-out is achieved near the
nucleation epoch and we therefore consider only the nucleation
temperature  ($T_{\mbox{\tiny{nucl}}}$). 

In principle, this condition can be translated to a bound on the order
parameter of the PT, or on the Higgs mass~\cite{Kajantie}.  
However, there are a number of approximations and nontrivial steps
involved in this procedure. 
The condition~(\ref{sphal2}) is on the sphaleron energy at the
temperature of bubble nucleation and it has to be related to the
vacuum expectation value of the Higgs field (\emph{vev}), at the
critical temperature ($T_c$). 
These two temperatures are not exactly the same since, though the
quantum tunneling phenomenon starts at $T_c$, initially the bubbles
are not large enough for their volume energy to overcome the surface
tension and they shrink. We have to wait for a lower temperature
$T_{\mbox{\tiny{nucl}}}$, when bubbles are large enough to
grow. Besides, $M_H$ is assumed to be equal to $M_W$ and there are a
number of poorly known prefactors involved. In spite of these
difficulties, a condition to avoid the sphaleron erasure is found
and at present generally accepted:  
\be
   (\phi / T)_{min} \simeq 1.0 - 1.5 
\label{washout}\, .
\ee
This bound represents a condition on the order of the PT, requiring a
remarkable jump in the Higgs field.  

On another hand, the order of the EWPT depends on the mass of all the
particles of the theory (the $SU(2)_L \times U(1)_Y$ SM) and in
particular on the Higgs boson mass M$_H$, which is at present not
known. We only have constraints on it: the current lower bound on the
Higgs mass from LEP~\cite{LEP} is:  $m_H \simgreat 114$ GeV. 

The effective potential for the Higgs field, at finite temperature,
can be written, including the radiative corrections from all the known
SM particles:  

\be
   V_{\mbox{\tiny{eff}}} \simeq -\frac{1}{2}(\mu^2-\alpha T^2)\phi^2 - T \delta \phi^3 +    \frac{1}{4}(\lambda - \delta \lambda_T) \phi^4
\label{V_eff}\, , 
\ee
where the coefficients depend on the masses of the heaviest particles,
on the temperature and on the \emph{vev} (for details, see
e.g.~\cite{Elmfors}). 
The cubic term in the effective potential is responsible for the
existence of the barrier between the two degenerate vacua at $T_c$
which makes the transition first order. 

From the effective potential, the value of $(\phi / T)$ can be
estimated at the critical value $T_c$, when the two minima of the
potential become degenerate: $\phi / T = 2\delta / (\lambda - \delta
\lambda_T)$. This is proportional to the inverse of the Higgs mass and
its maximum value is 0.55 for $m_H = 0$. The transition weakens with
increasing Higgs mass, a result that is basically in agreement with
lattice calculations for the EWPT in the standard
model~\cite{Kajantie}. 

These values for $(\phi / T)$ do not overlap with those for the
requirement~(\ref{washout}) to avoid the sphaleron wash-out.

\section{Hypermagnetic fields and phase transitions}\label{III}

For temperatures above the EWPT, the $SU(2)_L\times U(1)_Y$ symmetry
is restored, the magnetic fields correspond to the $U(1)_Y$ group
instead of to the $U(1)_{em}$ group and they are therefore properly
called \emph{hypermagnetic} fields.  The only fields able to propagate
for long 
distances are the Abelian vector modes that represent a magnetic
field. On the other hand, electric fields~\cite{LeBellac} as well as 
non-Abelian fields are screened due to the development of a
temperature dependent mass. 


The hypercharge field $B_Y$ contains a component of the vector field
Z, which becomes massive in the broken phase and is thus screened,
such as a magnetic field is in a superconductor. The presence of a
hypermagnetic field consequently introduces an extra contribution to
the pressure term in the symmetric phase, enhancing the difference in
free energies between the two phases, making the PT more strongly
first order. 
Recently, it has been shown~\cite{Elmfors},~\cite{Giovannini} and
~\cite{Kajantie2}, using quite different methods (perturbatively, at
tree level and at one loop, and non perturbatively, with lattice
calculations) that hypermagnetic fields strengthen the PT, although
the calculations differ somewhat on the level of strength reached and on
the viable range for the Higgs mass and the field
value. Ref.~\cite{Elmfors} concludes that for $B_{Yc} \geq 0.33 T_c^2$, the
bound~(\ref{washout}) is preserved. Lattice
calculations~\cite{Kajantie2} have shown that, even high magnetic
field values do not suffice to obtain a first-order transition for
$m_H \geq 80$ GeV.  

Unfortunately, in the presence of an external magnetic field, the
relation between the 
\emph{vev} and the sphaleron energy is altered and even if
condition~(\ref{washout}) is respected, condition~(\ref{sphal2}) may
not be fulfilled anymore. 
In fact, another aspect that needs to be considered is the effect of
the magnetic field on the height of the sphaleron barrier, through the
coupling with the sphaleron's dipole moment. Ref.~\cite{Comelli} has
found that in this case the barrier is lowered, facilitating the
transition between 
topologically inequivalent vacua. These calculations conclude that
there is no value of $B_Y$ that is enough to push the sphaleronic
transitions out of thermal equilibrium. (See also~\cite{Skalozub}). 

\section{Magnetic fields in the universe}\label{IV}

Magnetic fields seem to be pervading the entire universe. They have been
observed in galaxies, clusters, intracluster medium and high-redshift
objects~\cite{Kron}. Estimation of magnetic fields strengths --by
synchrotron emission and Faraday rotation-- require the independent
estimation or assumption of the local electron density and the spatial
structure of the field. Both quantities are reasonably known for our
galaxy, where the average field strength is measured to be $3 - 4 \mu
G$; various spiral galaxies in our neighborhood present fields
that are homogeneous over galactic sizes, with similar magnetic field
intensities~\cite{Kron, Beck}. At larger scales, only model dependent
upper limits can be established. These limits are also in the few $\mu
G$ range. In the intracluster medium, recent results detect the
presence of $\mu G$ magnetic fields~\cite{Eilek, Clarke}. For
intergalactic large scale fields (dissociated from any particular
galaxy or cluster), an upper bound of $10^{-9} G$ has been estimated,
adopting some reasonable values for the magnetic coherence
length~\cite{Kron}.  

The origin of these fields is nowadays unknown but it is widely believed
that two ingredients are needed for their generation: a mechanism for
creating the seed fields and a process for amplifying both their
amplitude and their coherence scale~\cite{Giovannini2, Reviews}.

Generation of the seed field (magnetogenesis) may be either primordial
or associated to the process of structure formation.  
In the early universe, which is the case of interest here, there are a
number of proposed mechanisms that could possibly generate primordial
magnetic fields.  Among the best suited are first order phase
transitions~\cite{Quash, Baym, Boyan}, which provide favorable
conditions for magnetogenesis such as charge separation, turbulence
and out-of-equilibrium conditions. Local charge 
separation, creating local currents, can be achieved through the high
pressure effect on the different equations of state of baryons and
leptons, behind the shock fronts which precede the expanding bubbles.  
Turbulent flow near the bubble walls is then expected to amplify
and freeze the seed field and when two shock fronts collide, turbulence
and vorticity --and hence magnetic fields-- can be generated to larger
scales. Other proposals include bubble wall collisions, which produce
phase gradients of the complex order parameter that act as a source for
gauge fields~\cite{Kibble}.  A low expansion velocity of the bubbles
wall then allows the magnetic flux generated in the intersection region to
penetrate the colliding bubbles. 

When interested in larger coherence scales, a plausible scenario is
inflation, where super-horizon scale fields are generated through the
amplification of quantum fluctuations of the gauge fields. This
process needs however a mechanism for breaking conformal invariance
of the electromagnetic field~\cite{Dolgov1}. Several possibilities
have been proposed, introducing non-minimal coupling of photons to
curvature~\cite{Turner}, to the dilaton/inflaton field~\cite{Ratra} or
to fermions~\cite{Prokopec}.  

The most promising way to distinguish between primordial
and protogalactic fields is searching for their imprint on the cosmic
microwave background radiation (CMBR). A homogeneous magnetic field would
spoil the universe isotropy, giving rise to a dipole anisotropy in the
background radiation; on this basis, COBE results set an upper bound on
the present equivalent field strength~\cite{Barrow} at the level of $10^
{-9}$G. On the other hand, primordial magnetic fields affect the wave
patterns generating fluctuations in the energy density, producing
distortions in the Planckian spectrum~\cite{Jedamzik} and on the
Doppler peaks~\cite{Adams}.  
Here the bounds depend on the coherence scale~\cite{Jedamzik,
Mack}. The polarization can also be affected by primordial magnetic
fields, through depolarization~\cite{Harari} and cross-correlations
between temperature and polarization anisotropies~\cite{Scann}. The
future CMBR satellite mission PLANCK may reach the required
sensitivity for the detection of these last signals.  

Although there is at present no conclusive evidence about the origin
of magnetic fields, their existence prior to the EWPT epoch cannot
certainly be ruled out. We will then work with primordial
hypermagnetic fields, homogeneous over bubble scales.   

\section{CP violating fermion scattering with hypermagnetic
fields}\label{V}

During the EWPT, the
properties of the bubble wall depend on the effective, finite temperature
Higgs potential. Under the assumption that the wall is thin and that
the phase transition happens when the energy densities of both phases
are degenerate, it is possible to find a one-dimensional analytical
solution for the Higgs field $\phi$ called the \emph{kink}. When
scattering is not affected by diffusion, the problem of fermion
reflection and transmission through the wall can be cast in terms of
solving the Dirac equation with a position dependent fermion mass,
proportional to the Higgs field~\cite{Ayala4}. In order to simplify
the discussion, let us consider a situation in which we
take the limit when the width of the wall approaches zero. In this
case, the kink solution becomes a step function $\Theta (z)$, where
$z$ is the coordinate along the direction of the phase
change~\cite{Ayala1}. Since the mass of the particles is dictated by
its coupling to the Higgs field, in our approximation, the former is
given by 
\be
   m(z)=m_0\Theta (z)\, .
   \label{step}
\ee   
In terms of Eq.~(\ref{step}), we can see that $z\leq 0$ represents
the region outside the bubble, that is the region in the symmetric
phase where particles are massless. Conversely, for $z\geq 0$, the system
is inside the bubble, that is in the broken phase and particles have
acquired a finite mass $m_0$.

In the presence of an external magnetic field, we need to consider
that fermion modes couple differently to the field in the broken and
the symmetric phases. Let us first look at the symmetric phase.

For $z\leq 0$, the coupling is chiral. Let 
\be
   \Psi_R&=&\frac{1}{2}\left(1 + \gamma_5\right)\Psi\nonumber\\
   \Psi_L&=&\frac{1}{2}\left(1 - \gamma_5\right)\Psi
   \label{chiralmodes}
\ee
represent, as usual, the right and left-handed chirality modes for the
spinor $\Psi$, respectively. Then, the equations of motion for these
modes, as derived from the electroweak interaction Lagrangian, are
\be
   (i\partial\dia\ -\ \frac{y_L}{2}g'A\dia\ )
   \Psi_L - m(z)\Psi_R &=& 0\nonumber\\
   (i\partial\dia\ -\ \frac{y_R}{2}g'A\dia\ )
   \Psi_R - m(z)\Psi_L &=& 0\, ,
   \label{diracsymm}
\ee 
where $y_{R,L}$ are the right and left-handed hypercharges
corresponding to the given fermion, respectively, $g'$ the
$U(1)_Y$ coupling constant and we take $A^\mu=(0,{\mathbf A})$
representing a, not as yet specified, four-vector potential having
non-zero components only for its spatial part, in the rest frame of
the wall. 

The set of Eqs.~(\ref{diracsymm}) can be written as a single equation
for the spinor $\Psi = \Psi_R + \Psi_L$ by adding up the former equations
\be
   \left\{ i\partial\dia\ - A\dia
   \left[\frac{y_R}{4}g'\left(1 + \gamma_5\right)
   +\frac{y_L}{4}g'\left(1 - \gamma_5\right)\right]
   - m(z)\right\}\Psi = 0\, .
   \label{diracsingle}
\ee   
Hereafter, we explicitly work in the chiral representation of the
gamma matrices 
\be
   \gamma^0=\!\left(\begin{array}{rr}
   0 & -I \\
   -I & 0 \end{array}\right)\
   \mbox{\boldmath $\gamma$}=\!\left(\begin{array}{rr}
   0 &  \mbox{\boldmath $\sigma$} \\
   \mbox{\boldmath $-\sigma$} & 0 \end{array}\right)\
   \gamma_5=\!\left(\begin{array}{rr}
   I & 0 \\
   0 & -I \end{array}\right)
   \label{gammaschiral}\, ,
\ee
and thus write Eq.~(\ref{diracsingle}) as
\be
   \Big\{i\partial\dia\ -\ {\mathcal G}A_\mu\gamma^\mu
   -m(z)\Big\}\Psi=0\, ,
   \label{diracsimple}
\ee
where we have introduced the matrix
\be
   {\mathcal G}=\left(\begin{array}{cc}
   \frac{y_L}{2}g'I & 0 \\
   0 & \frac{y_R}{2}g'I \end{array}\right)\, .
   \label{matA}
\ee
We now look at the corresponding equation in the broken symmetry
phase. For $z\geq 0$ the coupling of the fermion with the external 
field is through the electric charge $e$ and thus, the equation of motion
is simply the Dirac equation describing an electrically
charged fermion in a background magnetic field, namely,
\be
   \Big\{i\partial\dia\ -\ eA_\mu\gamma^\mu
   -m(z)\Big\}\Psi=0\, .
   \label{diracsimplezg0}
\ee
For definiteness, let us consider a constant magnetic field
${\mathbf B}=B\hat{z}$ pointing along the $\hat{z}$ direction. In this
case, the vector potential ${\mathbf A}$ can only have components
perpendicular to $\hat{z}$ and the solution to the above equations
factorize, namely 
\be
   \Phi (t,\mbox{\boldmath $x$})=\zeta (x,y)\Phi (t,z)\, .
   \label{facto}
\ee
We concentrate on the solution describing the motion of fermions
perpendicular to the wall, {\it i.e.}, along the $\hat{z}$
axis and thus effectively treating the problem as the motion of
fermions in one dimension. The case where the width of the wall is
allowed to become finite has been addressed in Ref.~\cite{Ayala2} and
we will briefly present below the results of this more realistic case; 
whereas the motion of the fermions in three dimensions has been given
a solution in Ref.~\cite{Ayala3}.
  
Equations~(\ref{diracsimple}) and~(\ref{diracsimplezg0}) can be solved
analytically. We look for the scattering states 
appropriate to describe the motion of fermions in the symmetric
and broken symmetry phases. For our purposes, these are fermions
incident toward and reflected from the wall in the symmetric
phase. There are two types of such solutions; those 
coupled with $y_L$ and those coupled with $y_R$. For an incident wave
coupled with $y_L$ ($y_R$), the fact that the differential equation
mixes up the solutions means that the reflected wave
will also include a component coupled with $y_R$ ($y_L$). In analogy,
the solution to Eq.~(\ref{diracsimplezg0}) is found by looking for the
scattering states appropriate for the description of transmitted
waves. The solutions are explicitly constructed in
Ref.~\cite{Ayala1} to where we refer the reader for details. For the
purposes of this work, we proceed to describe how to use these
solutions to construct the transmission and reflection probabilities.

In order to quantitatively describe the scattering of fermions, we
need to compute the corresponding reflection and transmission
coefficients. These are built from the reflected, transmitted and
incident currents of each type. Recall that for a given spinor wave
function $\Psi$, the current normal to the wall is given by
\be
   J=\Psi^\dagger\gamma^0\gamma^3\Psi\, .
   \label{current}
\ee
The reflection and transmission coefficients, $R$ and $T$,
are given as the ratios of
the reflected and transmitted currents, to the incident one,
respectively, projected along a unit vector normal to the
wall.

\begin{figure}[t] 
\centering
\includegraphics[height=4.5cm]{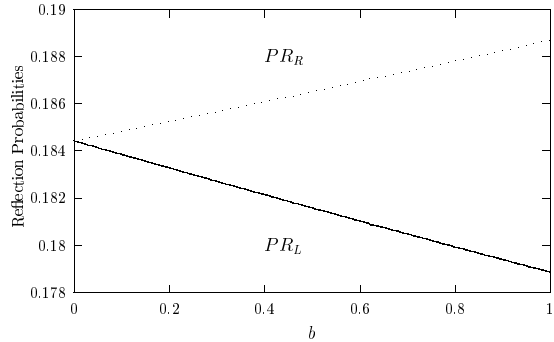}
\vspace{0.5cm}
\caption{Probabilities $PR_L$ and $PR_R$ as a function of the magnetic
field parametrized as $B=bT^2$ for $T=100$ GeV, $E=184$ GeV and a top
quark with a mass $m_0=175$ GeV, $y_R=4/3$, $y_L=1/3$. The value for
the U$(1)_Y$ coupling constant is taken as $g'=0.344$, corresponding
to the EWPT epoch.}
\label{fig1}
\end{figure}

The probabilities for finding a left or a right-handed
particle in the symmetric phase after reflection, $PR_L$, $PR_R$ are
given, respectively by
\be
   PR_L=R_{L\rightarrow L}+R_{R\rightarrow L}
\ee
\be
   PR_R=R_{L\rightarrow R}+R_{R\rightarrow R}\, ,
   \label{PR}
\ee
whereas the probabilities for finding a left or a right-handed
particle in the symmetry broken phase after transmission, $PT_L$, $PT_R$ are
given, respectively by
\be
   PT_L=T_{L\rightarrow L}+T_{R\rightarrow L}
\ee
\be
   PT_R=T_{L\rightarrow R}+T_{R\rightarrow R}\, .
   \label{PT}
\ee
Figure~\ref{fig1} shows the probabilities $PR_L$ and $PR_R$ as a
function of the magnetic field parametrized as $B=bT^2$ for a
temperature $T=100$ GeV, a fixed $E=184$ GeV and for a fermion taken
as the top quark with a mass $m_0=175$ GeV, $y_R=4/3$, $y_L=1/3$ and
for a value of $g'=0.344$, as appropriate for the EWPT epoch. Notice
that when $b\rightarrow 0$, these probabilities approach each other
and that the difference grows with increasing field strength. We have
considered the top quark since it is assumed to be the heaviest
particle in the broken phase, and hence to have the larger Yukawa
coupling. 


In the case that we allow for the vacuum expectation value of the
Higgs field and $m(z)$ to vary continuously through the wall, we work
in the thin wall regime with the kink solution: 
\be
   \varphi (x) = 1 + \tanh (x)\, ,
   \label{kink}
\ee
where the dimensionless position coordinate $x$ is proportional to $z$. 
In this case, we have to solve the same equation (\ref{diracsingle}),
but with a different $m(z)$ profile. We achieved this with a
combination of analytical and numerical methods and we report here
only the results for the reflection probabilities.  

Figure~2 shows the coefficients $R_{l\rightarrow r}$ and
$R_{r\rightarrow l}$ as a function of the magnetic field parameter $b
\equiv \left(\delta T / \sqrt {2 \lambda}\right)^{-2} B$, 
an energy parameter $\epsilon \equiv \left(\delta T / \sqrt {2
\lambda}\right)^{-1} E = 7.03$ (slightly larger than the height of
the barrier, in order to avoid the exponential damping of the
transmitted waves), and the other parameters as in the previous
case. Again, notice that when $b\rightarrow 0$, these coefficients
approach each other and that the difference grows with increasing
field strength. 
The results are in good agreement with the simplest case. This scheme
is more realistic then the former since the
height and width of the wall are typically related to each other in
such a way that it is not entirely realistic to vary one without
affecting the other.  

Though not explicitly worked out here, it is easy to convince
oneself that when considering the scattering in three dimensions, the
quantum mechanical motion of the fermion will include in general the
description of its velocity vector with a component perpendicular
to the field. In this case, due to the Lorentz force,
the particle circles around the field lines maintaining its
velocity along the direction of the field. The motion of the
particle is thus described as an overall displacement along the
field lines superimposed to a circular motion around these
lines~\cite{Ayala3}. These circles are labeled by the principal
quantum number. We see that if we have fermions that start off moving
by making a nonzero angle with the field lines, all of these
trajectories will result at the end in the same overall direction of
incidence. Also, since the fermion coupling with the external field is
through its spin, changing the direction of the field exchanges the
role of each spin component but since each chirality mode contains
both spin orientations, this does not affect the final probabilities
and thus the asymmetry is independent of the orientation of the field
with respect to the $\hat{z}$ axis.

It is interesting to notice that, with this mechanism, we are not
generating a net excess of one type of particle (left- or right-
handed) over the other; it is merely a segregation between the two
sides of the bubble wall. 

We also emphasize that, under the very general assumptions of CPT
invariance and unitarity, the total axial asymmetry (which includes
contributions both from particles and antiparticles) is quantified in
terms of the particle axial asymmetry. Let $\rho_i$ represent the
number density for species $i$. The net densities in left-handed
and right-handed axial charges are obtained by taking the differences
$\rho_L-\rho_{\bar{L}}$ and $\rho_R-\rho_{\bar{R}}$, respectively. It
is straightforward to show~\cite{Nelson} that CPT invariance and
unitarity imply that the above net densities are given by
\be
   \rho_L-\rho_{\bar{L}}&=&(f^s-f^b)
   (R_{r\rightarrow l} - R_{l\rightarrow r})\nonumber\\
   \rho_R-\rho_{\bar{R}}&=&(f^s-f^b)
   (R_{l\rightarrow r} - R_{r\rightarrow l})\, ,
   \label{net} 
\ee  
where $f^s$ and $f^b$ are the statistical distributions for
particles or antiparticles (since the chemical potentials are assumed
to be zero or small compared to the temperature, these distributions
are the same for particles or antiparticles) in the symmetric and the
broken symmetry phases, respectively. From Eq.~(\ref{net}), the
asymmetry in the axial charge density is finally given by
\be
   (\rho_L-\rho_{\bar{L}}) - (\rho_R-\rho_{\bar{R}})=
   2(f^s-f^b)(R_{r\rightarrow l} - R_{l\rightarrow r}).
   \label{final}
\ee

\begin{figure}[t] 
\vspace{-0.4cm}
{\centering\rotatebox{-90}{
\resizebox*{0.50\textwidth}{!}
{\includegraphics{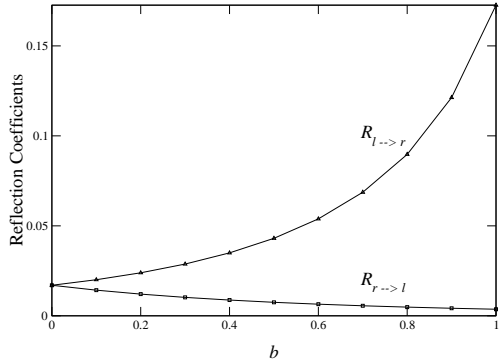}}}\par} 
\caption{Coefficients $R_{l\rightarrow r}$ and
$R_{r\rightarrow l}$, in the case of a smooth variation of $m(z)$, as
a function of the magnetic 
field parameter $b$ for an energy parameter $\epsilon=7.03$, $y_R=4/3$,
$y_L=1/3$. The value for the $U(1)_Y$ coupling constant is taken as
$g'=0.344$, corresponding to the EWPT epoch. The dots represent the
computed values.} 
\end{figure}

This asymmetry, built on either side of the
wall, is dissociated from non-conserving baryon number 
processes and can subsequently be converted to baryon number in the
symmetric phase where sphaleron induced transitions are taking
place with a large rate. This mechanism receives the name of 
\emph{non-local baryogenesis}~\cite{{CKN},{Nelson},{Dine},{Joyce}}
and, in the absence of the external field, it can only be realized in
extensions of the SM where a source of CP violation
is introduced ad hoc into a complex, space-dependent phase of
the Higgs field during the development of the EWPT~\cite{Torrente}. 

In our case, the relation of this axial asymmetry to CP violation is
understood as follows: recall that for instance, in the SM, CP is
violated in the quark sector through the mixing between different weak
interaction eigenstates to form states with definite mass. However, in
the present scenario, no such mixing occurs since we are concerned
only with the evolution of a single quark ({\it e.g.}, the top quark)
species. The relation is thus to be found in the dynamics of the
scattering process itself and becomes clear once we notice that this
can be thought of as describing the mixing of two levels, right- and
left-handed quarks coupled to an external hypermagnetic field. When
the two chirality modes interact only with the external field,
they evolve separately. It is only the scattering with the bubble wall
what allows a finite transition probability for one mode to become the
other. Since the modes are coupled differently to the external field, these
probabilities are different and give rise to the axial asymmetry. CP
is violated in the process because, though C is conserved, P is
violated and thus is CP.

\section{Summary and outlook}\label{VI}

In this work we have given a quantitative outline of the CP
violating scattering of fermions off (a simplified picture of) EWPT
bubbles in the presence of hypermagnetic fields. This scattering
produces an axial asymmetry built on either side of the bubble walls. The 
origin of this asymmetry is the chiral nature of the fermion coupling
to the hypermagnetic field in the symmetric phase. We have shown how
to compute reflection and transmission coefficients and also that
these differ for left and right-handed incident particles.

Primordial hypermagnetic fields thus provide with a much needed
ingredient, namely, additional CP violation, for the possible
generation of baryon number during the EWPT. A second ingredient, the
strengthening of the order of the phase transition and thus the
avoidance of the sphaleron bound seems at the moment a difficult
problem to surmount. Nonetheless, it is important to bear in mind that so
far, the calculations that provide insight into the effect of the
hypermagnetic fields on the order of the EWPT do not account for
the non-perturbative effects, cast in the language of resummation,
which are otherwise well known to play a very important role for the
dynamics of the phase transition in the absence of magnetic
fields. Much work is needed in this direction. This is for the future.

\section*{Acknowledgments}

GP is grateful to Instituto de Astronom\'{\i}a, UNAM, for its kind
hospitality during the development of this work. Support for this work
has been received in part by DGAPA-UNAM under 
PAPIIT grant number IN108001 and by CONACyT-M\'exico
under grant numbers 32279-E and 40025-F.

%
%
%
%

%

\begin{thebibliography}{99.}
%
%
%



\bibitem{Sakharov}
A. D. Sakharov, Pis'ma Zh. Eksp. Teor. Fiz. \textbf{5}, 32 (1967) [JETP
Lett. \textbf{5}, 24 (1967)].

\bibitem{Gavela}
M.B. Gavela, P. Hern\'andez, J. Orloff, and O. P\`ene,
Mod. Phys. Lett. A \textbf{9}, 795 (1994).

\bibitem{Kajantie}
K. Kajantie, M. Laine, K. Rummukainen, and M. Shaposnikov,
Nucl. Phys. B \textbf{466}, 189 (1996).

\bibitem{CKN}
A. G. Cohen, D. B. Kaplan and A. E. Nelson, Phys. Lett. B \textbf{263}, 86
(1991).

\bibitem{Carena}
M. Carena, M. Quiros and C. E. Wagner, Nucl. Phys. B \textbf{524}, 3 (1998).

\bibitem{Cline}
J. M. Cline, {\it Electroweak phase transition and baryogenesis},
hep-ph/0201286.

\bibitem{Giovannini}
M. Giovannini and M. E. Shaposhnikov, Phys. Rev. D \textbf{57}, 2186 (1998).

\bibitem{Elmfors}
P. Elmfors, K. Enqvist and K. Kainulainen, Phys. Lett. B \textbf{440},
269 (1998). 

\bibitem{Giovannini2}
M. Giovannini, {\it Primordial Magnetic Fields}, hep-ph/0208152.

\bibitem{Comelli}
D. Comelli, D. Grasso, M. Pietroni and A. Riotto, Phys. Lett. B
\textbf{458}, 304 
(1999). 

\bibitem{Carrington}
M. E. Carrington, Phys. Rev. D \textbf{45}, 2933 (1992).

\bibitem{Ayala1} 
A. Ayala, J. Besprosvany, G. Pallares and G. Piccinelli, Phys. Rev. D
\textbf{64}, 123529 (2001).  

\bibitem{Ayala2} 
A. Ayala, G. Piccinelli and G. Pallares, Phys. Rev. D \textbf{66},
103503 (2002). 

\bibitem{Ayala3} 
A. Ayala and J. Besprosvany, Nucl. Phys. B \textbf{651}, 211 (2003).

\bibitem{Nelson}
A. E. Nelson, D. B. Kaplan and A. G. Cohen, 
Nucl. Phys. B \textbf{373}, 453 (1992).

\bibitem{Trodden}
M. Trodden, Rev. Mod. Phys. \textbf{71}, 1463 (1999).

\bibitem{Riotto}
A. Riotto and M. Trodden, Ann. Rev. Nucl. Part. Sci. \textbf{49}, 35 (1999).

\bibitem{Rubakov}
V. A. Rubakov and M. E. Shaposhnivov, Usp.Fiz.Nauk \textbf{166}, 493 (1996), hep-ph/9603208.

\bibitem{Kolb}
E. W. Kolb and M. S. Turner: \textit{The Early Universe}, (Addison--Wesley Publishing Company 1990). 

\bibitem{Stecker}
F. W. Stecker: The matter-antimatter asymmetry of the universe. To be published in: {\it XIVth Rencontres de Blois 2002 on Matter-Antimatter Asymmetry}, ed. by J. Tran Thanh Van, hep-ph/0207323.

\bibitem{Dolgov}
A. D. Dolgov: Cosmological matter-antimatter asymmetry and antimatter in the universe. To be published in: \textit{XIVth Rencontres de Blois 2002 on Matter-Antimatter Asymmetry}, ed. by J. Tran Thanh Van, hep-ph/0211260.

\bibitem{Klink}
F. R. Klinkhamer and N. S. Manton, Phys. Rev. D \textbf{30}, 2212 (1984).

\bibitem{'tHooft}
G. 't Hooft, Phys. Rev. Lett. \textbf{37}, 8 (1976); Phys. Rev. D \textbf{14}, 3432 (1976). 

\bibitem{Benreu} 
For a recent review on the subject, see e.g. W. Bernreuther, Lect. Notes Phys. \textbf{591}, 237 (2002), hep-ph/0205279.

\bibitem{Kuzmin}
V. A. Kuzmin, V. A. Rubakov and M. E. Shaposhnikov, Phys. Lett. B \textbf{155}, 36 (1985).

\bibitem{Shap1}
M. E. Shaposhnikov, Nucl. Phys. B \textbf{287}, 757 (1987).

\bibitem{LEP}
K. Hagiwara et al., Phys. Rev. D \textbf{66}, 010001 (2002), http://pdg.lbl.gov/.

\bibitem{LeBellac}
M. Le Bellac: \textit{Thermal Field Theory}, (Cambridge University
Press, Cambridge 1996) pp 124--127.

\bibitem{Kajantie2}
K. Kajantie, M. Laine, J. Peisa, K. Rummukainen, and M. Shaposnikov,
Nucl. Phys. B \textbf{544}, 357 (1999).

\bibitem{Skalozub}
V. Skalozub and V. Demchik, {\it Can baryogenesis survive in the standard model due to strong hypermagnetic field?}, hep-ph/9909550. 

\bibitem{Kron}
The observational situation is discussed in P.P. Kronberg, Rep. Prog. Phys. \textbf{57}, 325 (1994); or more recently in J.-L. Han and R. Wielebinski, {\it Milestones in the Observations of Cosmic Magnetic Fields}, astro-ph/0209090.

\bibitem{Beck}
R. Beck, A. Brandenburg, D. Moss, A. Shukurov and D. Sokoloff, Annu. Rev. Astron. Astrophys. \textbf{34}, 155 (1996).

\bibitem{Eilek}
J. A. Eilek and F. N. Owen, Ap. J. \textbf{567}, 202 (2002).

\bibitem{Clarke}
T. E. Clarke, P. P. Kronberg and H. B\"ohringer, Ap. J. \textbf{547}, L111 (2001).

\bibitem{Reviews}
For reviews on the origin, evolution and some cosmological consequences of primordial magnetic fields see: K. Enqvist, Int. J. Mod. Phys. \textbf{D7}, 331 (1998); 
R. Maartens: Cosmological magnetic fields. In: \textit{International Conference on Gravitation and Cosmology}, Pramana \textbf{55}, 575 (2000) and references therein; 
D. Grasso and H.R. Rubinstein, Phys. Rep. \textbf{348}, 163 (2001).

\bibitem{Quash}
For quark-hadron PT: J. Quashnock, A. Loeb and D.N. Spergel, Ap. J. \textbf{344}, L49 (1989); 
B. Cheng and A.V. Olinto, Phys. Rev. D \textbf{50}, 2421 (1994); 
G. Sigl, A.V. Olinto and K. Jedamzik, Phys. Rev. D \textbf{55}, 4582 (1997).

\bibitem{Baym}
For EWPT: G. Baym, D. B\"odeker and L. McLerran, Phys. Rev. D \textbf{53}, 662 (1996). 

\bibitem{Boyan}
For a PT at a critical temperature larger than EW scale: D. Boyanovsky, H. J. de Vega and M. Simionato, {\it Large scale magnetogenesis from a non-equilibrium phase transition in the radiation dominated era}, hep-ph/0211022.

\bibitem{Kibble}
T. Vachaspati, Phys. Lett. B \textbf{265}, 258 (1991);  T. W. B. Kibble and A. Vilenkin, Phys. Rev. D \textbf{52}, 679 (1995); E.J. Copeland, P. M. Saffin and O. T\"ornqvist, Phys. Rev. D \textbf{61}, 105005 (2000).

\bibitem{Dolgov1}
For an overview of the subject, see e.g., A. D. Dolgov, {\it Generation of magnetic fields in cosmology}, hep-ph/0110293.

\bibitem{Turner}
M. S. Turner and L. M. Widrow, Phys. Rev. D \textbf{37}, 2743 (1988).

\bibitem{Ratra}
B. Ratra, Ap. J. \textbf{391}, L1 (1992); M. Gasperini, M. Giovannini and G. Veneziano, Phys. Rev. Lett. \textbf{75}, 3796 (1995); D. Lemoine and M. Lemoine, Phys. Rev. D \textbf{52}, 1955 (1995). 

\bibitem{Prokopec}
T. Prokopec, {\it Cosmological magnetic fields from photon coupling to fermions and bosons in inflation}, astro-ph/0106247.

\bibitem{Barrow}
J.D. Barrow, P. Ferreira and J. Silk, Phys. Rev. Lett. \textbf{78}, 3610 (1997).

\bibitem{Jedamzik}
K. Jedamzik, V. Katalini\'c and A. V. Olinto, Phys. Rev. Lett. \textbf{85}, 700 (2000).

\bibitem{Adams}
J. Adams, U. H. Danielsson, D. Grasso and H. rubinstein, Phys. Lett. B \textbf{388}, 253 (1996).

\bibitem{Mack} 
For the effect of stochastic magnetic fields, see A. Mack, T. Kahniashvili and A. Kosowsky, Phys.Rev. D \textbf{65}, 123004 (2002), and references therein.

\bibitem{Harari}
D. D. Harari, J. D. Hayward and M. Zaldarriaga, Phys. Rev. D \textbf{55}, 1841 (1997);
M Giovannini, Phys. Rev. D \textbf{56}, 3198 (1997).

\bibitem{Scann} 
A. Kosowsky and A. Loeb, Ap. J. \textbf{469}, 1 (1996);
E. S. Scannapieco and P.G. Ferreira, Phys. Rev. D \textbf{56}, 7493 (1997).

\bibitem{Ayala4}
A. Ayala, J. Jalilian-Marian, L. McLerran and A. P. Vischer,
Phys. Rev. D \textbf{49}, 5559 (1994).

\bibitem{Dine}
M. Dine, O. Lechtenfield, B. Sakita, W. Fischel and J. Polchinski,
Nucl. Phys. B \textbf{342}, 381 (1990).

\bibitem{Joyce}
M. Joyce, T. Prokopec and N. Turok, Phys. Lett. B \textbf{338}, 269 (1994).

\bibitem{Torrente}
E. Torrente-Lujan, Phys. Rev. D \textbf{60}, 085003 (1999). 


\end{thebibliography}
%



\printindex
\end{document}